\documentclass[12pt]{iopart}
\usepackage{iopams}
\usepackage{graphicx}
\usepackage{xcolor}

\begin{document}

\title[]{Topological Devil's staircase in atomic two-leg ladders}

\author{S Barbarino$^{1,2}$, D Rossini$^3$, M Rizzi$^4$, R Fazio$^{5,6}$, G E Santoro$^{1,7,5}$, M Dalmonte$^{1,5}$}
\address{$^{1}$ Scuola Internazionale Studi Superiori Avanzati (SISSA), Via Bonomea 265, 34136 Trieste, Italy}
\address{$^{2}$ Institute of Theoretical Physics, Technische Universit\"at Dresden, 01062 Dresden, Germany}
\address{$^{3}$ Dipartimento di Fisica, Universit\`a di Pisa and INFN, Largo Pontecorvo 3, 56127 Pisa, Italy}
\address{$^{4}$ Institute f\"ur Physik, Johannes Gutenberg-Universit\"at, 55128 Mainz, Germany}
\address{$^{5}$ Institute of Quantum Control (PGI-8), Forschungszentrum J\"ulich, D-52425 J\"ulich, Germany}
\address{$^{6}$ Institute for Theoretical Physics, University of Cologne, D-50937 K\"oln, Germany}
\address{$^{7}$ International Centre for Theoretical Physics (ICTP), P.O. Box 586, 34014 Trieste, Italy}
\address{$^{8}$ NEST, Scuola Normale Superiore \& Istituto Nanoscienze-CNR, 56126 Pisa, Italy}
\address{$^{9}$ CNR-IOM Democritos National Simulation Center, Via Bonomea 265, 34136 Trieste, Italy}

\submitto{\NJP}


\begin{abstract}
 We show that a hierarchy of topological phases in one dimension ---a topological Devil's staircase--- can emerge at fractional filling fractions in interacting systems, whose single-particle band structure describes a topological or a crystalline topological insulator.
Focusing on a specific example in the BDI class, we present a field-theoretical argument based on bosonization that indicates how the system, as a function of the filling fraction, hosts a series of density waves. Subsequently, based on a numerical investigation of the {{low-lying energy spectrum}}, Wilczek-Zee phases, and entanglement spectra, we show that {{they are symmetry protected topological phases.}} In sharp contrast to the non-interacting limit, these topological density waves do not follow the bulk-edge correspondence, as their edge modes are gapped. We then discuss how these results are immediately applicable to models in the AIII class, and to crystalline topological insulators protected by inversion symmetry. Our findings are immediately relevant to cold atom experiments with alkaline-earth atoms in optical lattices, where the band structure properties we exploit have been recently realized.
\end{abstract}

\maketitle

\section{Introduction}

In the last few years, a series of remarkable experiments has demonstrated how cold atomic gases in optical lattices can realize topological band structures~\cite{Altland97,Schnyder08,Kitaev09,Hasan_10,Qi_11,Ludwig15,Wen_17} with a high degree of accuracy and tunability~\cite{Cooper_18,Sengstock13,Aidelsburger15,Kennedy15,Tai17}. 
In the context of one-dimensional (1D) systems, ladders pierced by synthetic gauge fields~\cite{Orignac2001,Dhar2013,Tokuno2014,Greschner2016,Kolley2015,Piraud2015,Barbarino15,Barbarino16,Taddia16,orignac17,citro18}
 have been experimentally shown to display a plethora of phenomena, including chiral currents~\cite{Atala14} and edge modes akin to the two-dimensional Hall effect~\cite{Wen_17}, accompanied with the long-predicted - but hard to directly observe - skipping orbits~\cite{Mancini15,Stuhl15}. 
While such phenomena have required relatively simple microscopic Hamiltonians apt to describe electrons in a magnetic field~\cite{Jaksch03}, the flexibility demonstrated in very recent settings utilizing alkaline-earth-like atoms~\cite{cazalilla15,Livi16,Kolkowitz17, Kang18b,Kang18,Wall16} has shown how a new class of model Hamiltonians - where nearest neighbor couplings on multi-leg ladders can be engineered almost independently one from the other - is well within experimental reach. Remarkably, these works have not only demonstrated the capability of realizing spin-orbit couplings utilizing clock transitions~\cite{Livi16,Kolkowitz17}, but also the observation of band structures where topology is tied to inversion symmetry~\cite{Kang18b,Barbarino18}, a playground for crystalline topological insulators~\cite{Hughes_11, Chiu_13, Chiu_16}. 
A natural question along these lines is whether these new recently developed setups offer novel opportunities for the observation of intrinsically interacting topological phases - e.g., symmetry-protected topological phases which appear at fractional filling fractions.

In this work, we show how, starting from experimentally realized microscopic Hamiltonians~\cite{Kang18b}, interactions can generically stabilize novel topological phases in regimes where single-particle Hamiltonians cannot host any. We consider a 1D ladder with two internal spin states, supporting a topological phase at integer filling, and we show that, when the particle filling is reduced to a fractional value, repulsive interactions can stabilize a hierarchy of unconventional topological gapped phases, namely a topological Devil's staircase~\cite{Hubbard_78, Pokrovsky_78}. Such topological density-wave phases are characterized by a well-defined topological number, the Wilczek-Zee phase~\cite{Wilczek_84, Resta_94}, thus signaling that the topological properties of the non-interacting bands are inherited at fractional fillings in the presence of interactions. These gapped states present a degenerate entanglement spectrum~\cite{Fidkowski10,Pollmann10,Berg_11} and, in some regimes, an unconventional edge physics without zero-energy modes.

The appearance of these fractional topological phases is reminiscent of the quantum Hall physics, where a similar transition from the integer to the fractional regime is observed when interactions are considered. Owing to the 1D context, here the main difference is that all the phases are symmetry-protected topological phases, as true topological order cannot take place. We note that, for specific filling fractions, our results are closely related to other topological density waves found in single-band models~\cite{Guo_12, Budich_13}. The mere existence of a full class of topological density waves is surprising in view of the fact that, typically, non-interacting topological phases at integer filling appearing in the context of one-dimensional systems with two internal spin states such as the Su-Schrieffer-Hegger model~\cite{Su_79} or Creutz ladders~\cite{Creutz_99} are robust against weak interactions only, and disappear~\cite{Guo_11, Junemann_17} in the strongly interacting regime\footnote{This last fact is instead not surprising, once one realizes that, in the presence of a strongly repulsive Hubbard interaction term, such kind of models with nearest-neighbor hopping terms can be mapped onto topologically trivial spin-$1/2$ XYZ models~\cite{Essler_05}.}.

We illustrated the appearance of such phases by studying a model Hamiltonian description for the BDI~\cite{Su_79,Creutz_99,Guo_11} and AIII symmetry classes~\cite{Velasco_17,Junemann_17} of the Altland-Zirnbauer classification (AZc)~\cite{Altland97,Ludwig15}, and a crystalline topological insulator case of a 1D model supporting a spatial inversion symmetry protected topological phase at filling one~\cite{Hughes_11, Chiu_13, Chiu_16}. Our results are general in the sense that these fractional phases can be potentially observed in all symmetry classes of the AZc which can be realized in a two-leg ladder. Our work is complementary to recent approaches investigating interaction induced fractional topological insulators~\cite{carr2006,stoudenmire2011,huang2013,kraus2013,Petrescu16,Tovmasyan16,Strinati17,santos2018,Rachel18}, which typically focus on specific case scenarios that accurately mimic the edge physics of quantum Hall states or extend topological superconductivity at finite interaction strength.

From an experimental perspective, the models we investigate are immediately relevant to cold gases experiments. In particular, recent implementations using alkaline-earth-like atoms such as Yb~\cite{Mancini15, Livi16,Kang18} and Sr~\cite{Kolkowitz17} have demonstrated an ample degree of flexibility in tuning parameters (including static gauge fields) in two-leg ladders, exploiting the concept of synthetic dimension~\cite{Boada12, Celi14}. Most importantly, the single particle Hamiltonian we discuss below has been realized in a $^{173}$Yb gas, see Ref.~\cite{Kang18}, and similar schemes shall be applicable to $^{87}$Sr gases as well.

This paper is organized as follows.
In Sec.~\ref{model} we present the model and discuss its fundamental symmetries. 
{{Sections~\ref{BDI results} and~\ref{sec:DevilStair} contain our main results. 
In particular, in Sec.~\ref{BDI results}, we consider models belonging to the BDI and AIII symmetry classes and we show the appearance of a topological fractional phase at filling  $\nu=1/2$  which can be viewed as a precursor of the topological Devil's staircase. 
We discuss the topological properties of this phase: the Wilczek-Zee phase, and the entanglement spectrum in the ground-state manifold, using numerical methods as Lanczos-based exact diagonalization~\cite{Lanczos_50} and density-matrix renormalization group (DMRG)~\cite{White92,Schollwock_05} simulations.
Finally, we discuss how this topological phase supports edge modes, which, while not at zero energy, can still be diagnosed by simple correlation functions.
Then, in Sec.~\ref{sec:DevilStair}, by means of a bosonization approach we discuss the appearance of the topological Devil's staircase at lower fillings and we explicitly address the filling $\nu=1/3$ case.  Finally, in Sec.~\ref{inversion_results} we generalize our results to the case of a crystalline topological insulator.
Our conclusions are drawn in Sec.~\ref{conclusions}.}}

\section{Model and symmetries}
\label{model}

Let us start by introducing the Hamiltonian we are going to focus on. 
For the sake of clarity, we also review the main definitions of time-reversal, particle-hole, and chiral symmetry in the language of second quantization, which is best suited to the case of interacting systems.

\subsection{Model Hamiltonian}
We consider a 1D chain with $L$ sites along the physical dimension. These are populated by fermionic particles described by the
canonical operators $\hat c^{(\dagger)}_{j,\sigma}$, annihilating (creating) a fermion at site $j=1,\dots, L$, with two internal degrees of freedom labeled by $\sigma = \, \uparrow, \downarrow$ (resp. $+1, -1$).

\begin{figure}
\centering
  \includegraphics[width=0.65\textwidth]{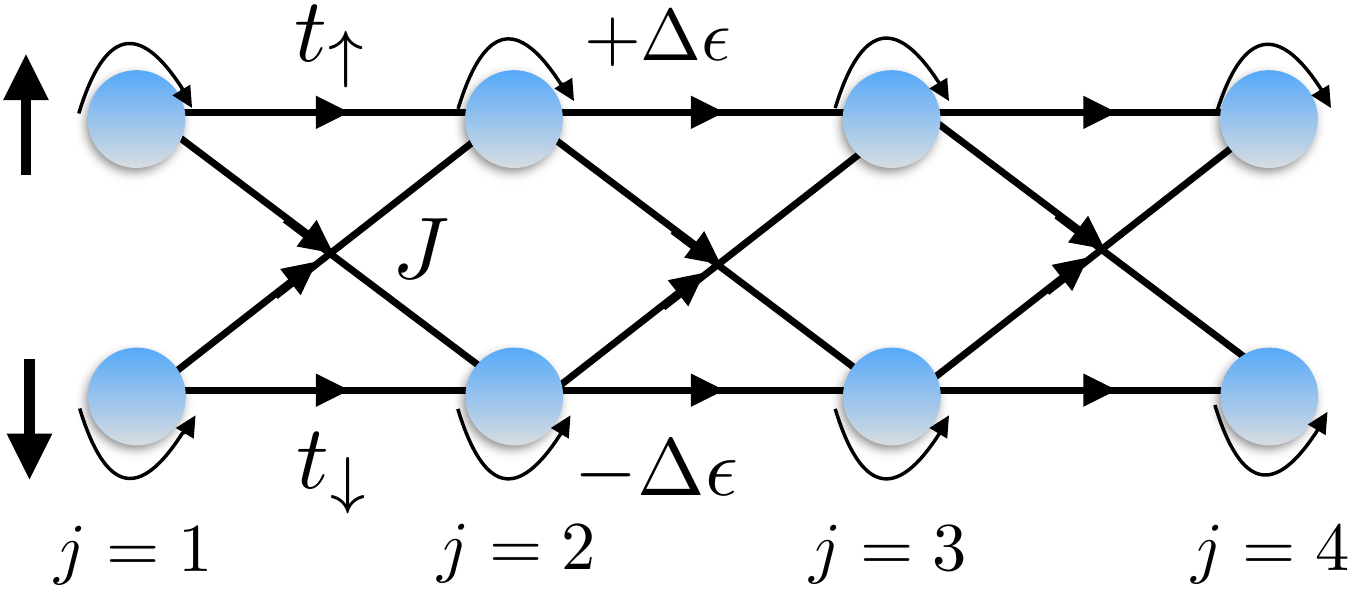}
  \caption{Schematic representation of a 1D chain with two internal spin states, $\uparrow$ and $\downarrow$. {{Curved arrows at the dots represent 
  the species-dependent chemical potential $\Delta \epsilon$ which induces an imbalance between spin up and spin down particles.
   As discussed in Table~\ref{fig:table}, by properly tuning the parameters $t_\sigma$ and $J$ in Eq.~(\ref{H0}),
    it is possible to realize non-interacting topological phases belonging to the symmetry classes BDI and AIII of the AZc 
    and an inversion symmetry protected topological phase supporting a non-trivial phase at integer filling (crystalline topological insulator).   }}}
  \label{fig:model}
\end{figure}

The single-particle physics discussed in this work is fully captured by the Hamiltonian [see Fig.~\ref{fig:model}]
\begin{equation}
  \hat H_0=\hat H_{nn}+\hat H_{\Delta \epsilon} ,
  \label{H}
\end{equation}
where 
\begin{equation}
  \hat H_{\Delta \epsilon}=\Delta \epsilon \sum_{\sigma, \,j} \sigma \, \hat c^\dagger_{j,\sigma} \hat c_{j,\sigma} 
\end{equation}
is a species-dependent chemical potential which induces an imbalance between spin up and spin down particles. The Hamiltonian term
\begin{equation}
  \hat H_{nn} = \sum_{\sigma, \, j}  \left( t_\sigma \, \hat c^\dagger_{j+1,\sigma} \hat c_{j,\sigma} + 
  J\, \hat c^\dagger_{j+1,\sigma} \hat c_{j,-\sigma}  \right) +\mathrm{H.c.}
  \label{H0}
\end{equation}
describes spin-preserving and spin-flipping nearest-neighbor hoppings; the parameters $t_\sigma$ and $J$ can be tuned according to the prescriptions of Table~\ref{fig:table}. 

Crucial for the following discussion is the presence of repulsive density-density interaction  terms
\begin{equation}
  \hat H_{{\rm int}}=U \sum_j \hat n_{j,\uparrow} \hat n_{j,\downarrow} + {\sum_p V_p \sum_j  \hat n_{j} \hat n_{j+p}} ,
\label{Hint}
\end{equation}
where $U>0$ and {$V_p>0$ with $p=1,\, 2,\dots$}; $\hat n_{j,\sigma}=\hat c^\dagger_{j,\sigma} \hat c_{j,\sigma}$ is the density operator and {$\hat n_{j} = \sum_\sigma \hat n_{j,\sigma}$ }. 
The full Hamiltonian is thus defined by
\begin{equation}
  \hat H = \hat H_0+ \hat H_{{\rm int}} = \hat H_{nn}+\hat H_{\Delta \epsilon} + \hat H_{{\rm int}} .
  \label{eq:HamFull}
\end{equation}
{{The model we discuss  has already been experimentally realized in Ref.~\cite{Kang18}.
We refer to this work for specific details on the experimentally achievable parameter regimes.}}
\newline
Hereafter we will set $\hbar=1$ and express all energy scales in units of the hopping term $t$.
In order to understand the symmetry class to which $\hat H$ belongs,
it is first convenient to recall the three symmetries classifying the ten classes of the AZc.
Then, we also discuss the inversion symmetry operator.

\begin{table}
\centering
  \includegraphics[width=0.65\textwidth]{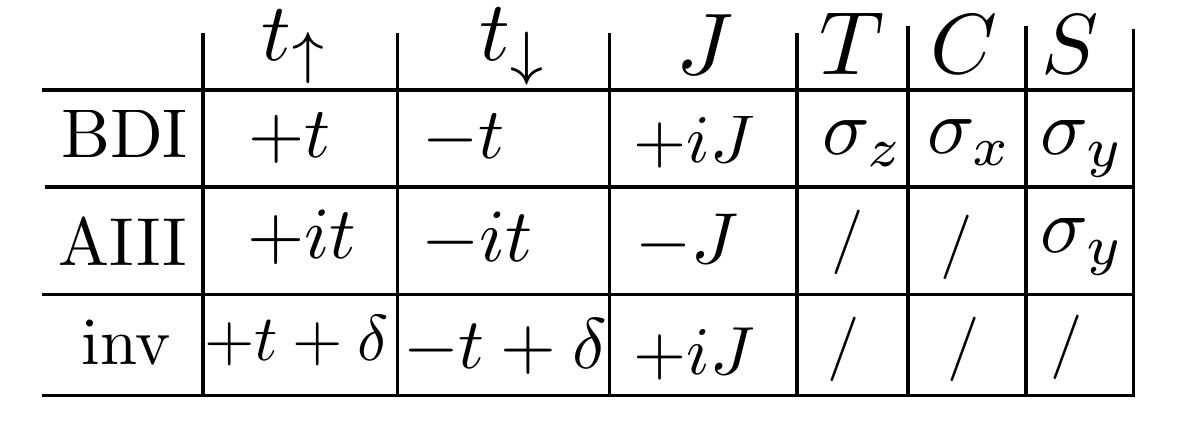}
  \caption{By properly tuning the parameters $t_\sigma$ and $J$ in Eq.~(\ref{H0}),
    it is possible to realize non-interacting topological phases belonging to the symmetry classes BDI and AIII of the AZc 
    and an inversion symmetry protected topological phase supporting a non-trivial phase at integer filling (crystalline topological insulator). 
    A topological phase belonging to the BDI symmetry class is endowed with a  time-reversal $T$, a particle-hole $C$, and a chiral $S$ symmetry; a topological phase in the AIII symmetry class is endowed with a chiral symmetry only; see also Eqs.~(\ref{timereversal})-(\ref{chiral}). }
  \label{fig:table}
\end{table}

\subsection{Fundamental symmetries}

The symmetries playing a crucial role in the AZc are the time-reversal $\hat \mathcal{ T}$,
the particle-hole $\hat \mathcal{ C}$, and the chiral $\hat \mathcal{ S}$ symmetry.
Their action on the fermionic operators $\hat c_{j,\sigma}$ reads~\cite{Ludwig15}: 
  \begin{eqnarray}
& \hat \mathcal{ T} \hat c_{j,\sigma}  \hat \mathcal{ T}^{-1} = T^*_{\sigma',\sigma} \hat c_{j,\sigma'}
 \hspace{0.8cm} 
\hat \mathcal{ T} \hat c^\dagger_{j,\sigma}  \hat \mathcal{ T}^{-1} = \hat c^\dagger_{j,\sigma'} T_{\sigma',\sigma} \label{timereversal} \\
& \hat \mathcal{ C} \hat c_{j,\sigma}  \hat \mathcal{ C}^{-1} = C_{\sigma',\sigma} \hat c^\dagger_{j,\sigma'}
 \hspace{0.95cm} 
 \hat \mathcal{ C} \hat c^\dagger_{j,\sigma}  \hat \mathcal{ C}^{-1} = \hat c_{j,\sigma'} C^*_{\sigma',\sigma}\label{particlehole} \\
& \hat \mathcal{ S} \,\hat c_{j,\sigma}\,  \hat \mathcal{ S}^{-1} = S_{\sigma',\sigma} \hat c^\dagger_{j,\sigma'}
 \hspace{0.8cm} 
 \hat \mathcal{ S} \hat c^\dagger_{j,\sigma}  \hat \mathcal{ S}^{-1} = \hat c_{j,\sigma'} S^*_{\sigma',\sigma} \label{chiral}
\end{eqnarray}
where $T$, $C$, and $S$ are $2\times 2$ unitary matrices satisfying $T T^*= C C^*= \pm \sigma_0$
($\sigma_0$ being the $2\times 2$ identity matrix) and $S= T C^*$, up to an arbitrary phase factor such that $S^2=\sigma_0$.
Furthermore, $\hat \mathcal{ T} $ and $\hat \mathcal{ S} $ are anti-unitary
[i.e., $\hat \mathcal{ T} (+i) \hat \mathcal{ T}^{-1} = \hat \mathcal{ S} \,(+i)\, \hat \mathcal{ S}^{-1} = -i$], 
while $\hat \mathcal{ C}$ is unitary [i.e., $\hat \mathcal{ C} (+i) \hat \mathcal{ C}^{-1} = +i$].
We also introduce the unitary inversion symmetry operator $\hat \mathcal{ I}$, which acts as~\cite{Chiu_13}: 
\begin{equation}
\hat \mathcal{ I} \,\hat c_{j,\sigma}\,  \hat \mathcal{ I}^{-1} = I_{\sigma',\sigma} \hat c_{-j,\sigma'}
 \hspace{0.6cm} 
 \hat \mathcal{ I} \, \hat c^\dagger_{j,\sigma} \, \hat \mathcal{ I}^{-1} = \hat c^\dagger_{-j,\sigma'} I^*_{\sigma',\sigma} \label{I}
\end{equation}
where $I$ is again a unitary $2\times 2$ matrix. 
The Hamiltonian $\hat H$ in Eq.~(\ref{eq:HamFull}) is invariant under a symmetry $\hat \mathcal{ M}$,
with $\hat \mathcal{ M} = \hat \mathcal{ T}, \; \hat \mathcal{ C}, \; \hat \mathcal{ S}$, if and only if
\begin{equation}
  \hat \mathcal{ M} \, \hat H \, \hat \mathcal{ M}^{-1} = \hat H.
\end{equation}
Switching off the interaction term,
the single-particle Hamiltonian~(\ref{H}) can be conveniently rewritten as:
{\begin{equation}
  \hat H_0 = \sum_k \hat C^\dagger_{k} \, H_0(k) \, \hat C_{k} ,
\end{equation}
with $\hat C^\dagger_{k}= (\hat c^\dagger_{k,\uparrow} \; \hat c^\dagger_{k,\downarrow})$} by means of the momentum-space operators 
\begin{equation}
  \hat c_{k,\sigma} =\frac{1}{\sqrt{L}} \sum_j e^{-ikj}\hat c_{j,\sigma} \,, \hspace{1cm} \; \mathrm{with} \;\; k \in [-\pi,\pi)\,.
\end{equation}
Then the requirements~(\ref{timereversal}-\ref{I}) lead  to the more familiar ones~\cite{Ludwig15}:
  \begin{eqnarray}
\hat \mathcal{ T} \, \hat H_0 \, \hat \mathcal{ T}^{-1} = \hat H_0 \quad & \rightarrow & \quad T \, H_0^*(k) \, T^\dagger=H_0(-k) \\
\hat \mathcal{ C} \, \hat H_0 \, \hat \mathcal{ C}^{-1} = \hat H_0 \quad & \rightarrow & \quad C \, H_0^*(k) \, C^\dagger=-H_0(-k) \qquad\\
\hat \mathcal{ S} \, \hat H_0 \, \hat \mathcal{ S}^{-1} = \hat H_0 \quad & \rightarrow & \quad S \, H_0(k) \, S^\dagger=-H_0(k) \\
\hat \mathcal{ I} \, \hat H_0 \, \hat \mathcal{ I}^{-1} = \hat H_0 \quad & \rightarrow & \quad I \, H_0(k) \, I^\dagger=H_0(-k) \,. 
\end{eqnarray}
According to the AZc, in 1D only five symmetry classes (BDI, AIII, D, CII, and DIII) can support a topological phase (assuming no spatial symmetry). 
In the next section we will consider interacting topological models whose single-particle Hamiltonians belong to the 
symmetry classes BDI~\cite{Guo_11} and AIII~\cite{Junemann_17, Velasco_17} and which can be realized 
in two-leg ladders with nearest-neighbor couplings, by properly tuning the coefficients $t_\sigma$ and $J$ in the Hamiltonian term $\hat H_{nn}$ of Eq.~(\ref{H0}), according to the prescriptions of Table~\ref{fig:table}.
On the other hand, CII and DIII models require ladders with a higher number of legs, or two-leg ladders in the presence of next-nearest-neighbor hopping terms, and will not be considered here. 

At integer filling, $\nu=N/L=1$ (where $N$ is the number of fermions), models in Refs.~\cite{Guo_11,Junemann_17, Velasco_17} can exhibit a topological phase characterized by the presence of exponentially localized zero-energy edge modes in the non-interacting spectrum, a quantized Zak phase, and a doubly degenerate entanglement spectrum~\cite{Fidkowski10,Pollmann10}.
Conversely, in the present context we are interested in investigating the topological properties when the particle filling is fractional, i.e. $\nu=1/q$, with $q>1$ integer. 
We will also consider ladders supporting a crystalline topological phase protected by the spatial inversion symmetry, which cannot be understood in terms of the standard AZc (Sec.~\ref{inversion_results}).

\section{Topological phases emerging due to interactions at fractional fillings in BDI and AIII band structures}
\label{BDI results}

Firstly we focus on two-leg ladder whose non-interacting Hamiltonian is in the BDI symmetry class.
In this case, the various parameters are fixed [see Table~\ref{fig:table}], and the resulting Hamiltonian of Eq.~(\ref{eq:HamFull}) reads:
\begin{equation}
\hat H = \; \sum_{\sigma, \, j} \left( \sigma t \, \hat c^\dagger_{j+1,\sigma} \hat c_{j,\sigma} + 
iJ\, \hat c^\dagger_{j+1,\sigma} \hat c_{j,-\sigma} + \mathrm{H.c.} \right) + \, \Delta \epsilon \sum_{\sigma, \, j} \sigma \, \hat c^\dagger_{j,\sigma} \hat c_{j,\sigma} + \hat H_{\rm int}\,.
\label{HBDI}
\end{equation}
Using Eqs.~(\ref{timereversal}-\ref{chiral}), { {and the matrices defined in Table~\ref{fig:table}}}, it is easy to observe that $\hat \mathcal{ T} \, \hat H \, \hat \mathcal{ T}^{-1} = \hat H$, while 
$\hat \mathcal{ C} \; \hat H \; \hat \mathcal{ C}^{-1} = \hat H$ and $\hat \mathcal{ S} \; \hat H \; \hat \mathcal{ S}^{-1} = \hat H$ up to constant terms, $UL + 4\sum_pV_p$, and trivial chemical potential terms $(-U-2\sum_p V_p)\sum_j \hat n_{j} $.
In the following we will focus on fractional fillings $\nu=1/q$, and consider repulsive interactions.
Our results for the BDI symmetry class are immediately applicable to the model in the AIII class, which can be obtained from to latter via the unitary transformation {{also known as Kawamoto-Smit rotation in the context of Lattice Field Theories \cite{Kawamoto}}}
$\hat \mathcal{ U}_j \, \hat c_{j,\sigma} \, \hat \mathcal{ U}_j^{-1} = e^{i \frac{\pi}{2}j} \hat c_{j,\sigma}$ [see again Table~\ref{fig:table}]

\subsection{Effective lowest-band Hamiltonian}

The single-particle contributions of the Hamiltonian~(\ref{HBDI}), assuming periodic boundary conditions (PBC), can be diagonalized as
\begin{equation}
  \hat H_0 = \sum_{k}\sum_{\eta} \eta \,E(k) \hat d^\dagger_{k,\eta}\, \hat d_{k, \eta} \, ,
\end{equation}
with $\eta= \pm 1$, $k \in [-\pi,\pi)$ and 
\begin{equation}
  E(k)=\sqrt{(2  t \cos k +\Delta \epsilon)^2 +4 J^2 \sin^2 k}
\end{equation}
by means of the unitary transformation $R_k=\exp{[i  \theta_k \sigma_y /2] }$ {such that $R^\dagger_k H_0(k) R_k=E(k) \sigma_z$ } with 
\begin{equation}
  \sin \theta_k= 2J \frac{\sin k}{E(k)}\;, \;\;\;\;\;\;\;\;\;\;\;\;
  \cos \theta_k= \frac{2t \cos k+\Delta \epsilon}{E(k)} \; ;
\end{equation}
we stress here that $R_k$ is defined up to an arbitrary complex phase $e^{i\varphi_k}$.    
Consequently, the operators $\hat d_{k,+1}$ and $\hat d_{k,-1}$ can be related to the original ones as
\begin{equation}
\cases{
\hat d_{k,-1}= e^{i\varphi_k}\left(\cos \frac{\theta_k}{2} \hat c_{k,\uparrow} + \sin \frac{\theta_k}{2} \hat c_{k,\downarrow} \right)\\
\hat d_{k,+1}=e^{i\varphi_k}\left( -\sin \frac{\theta_k}{2} \hat c_{k,\uparrow} +\cos \frac{\theta_k}{2} \hat c_{k,\downarrow} \right) \,.
}
\label{unitary}
\end{equation}
\begin{figure}
  \includegraphics[width=0.95\columnwidth]{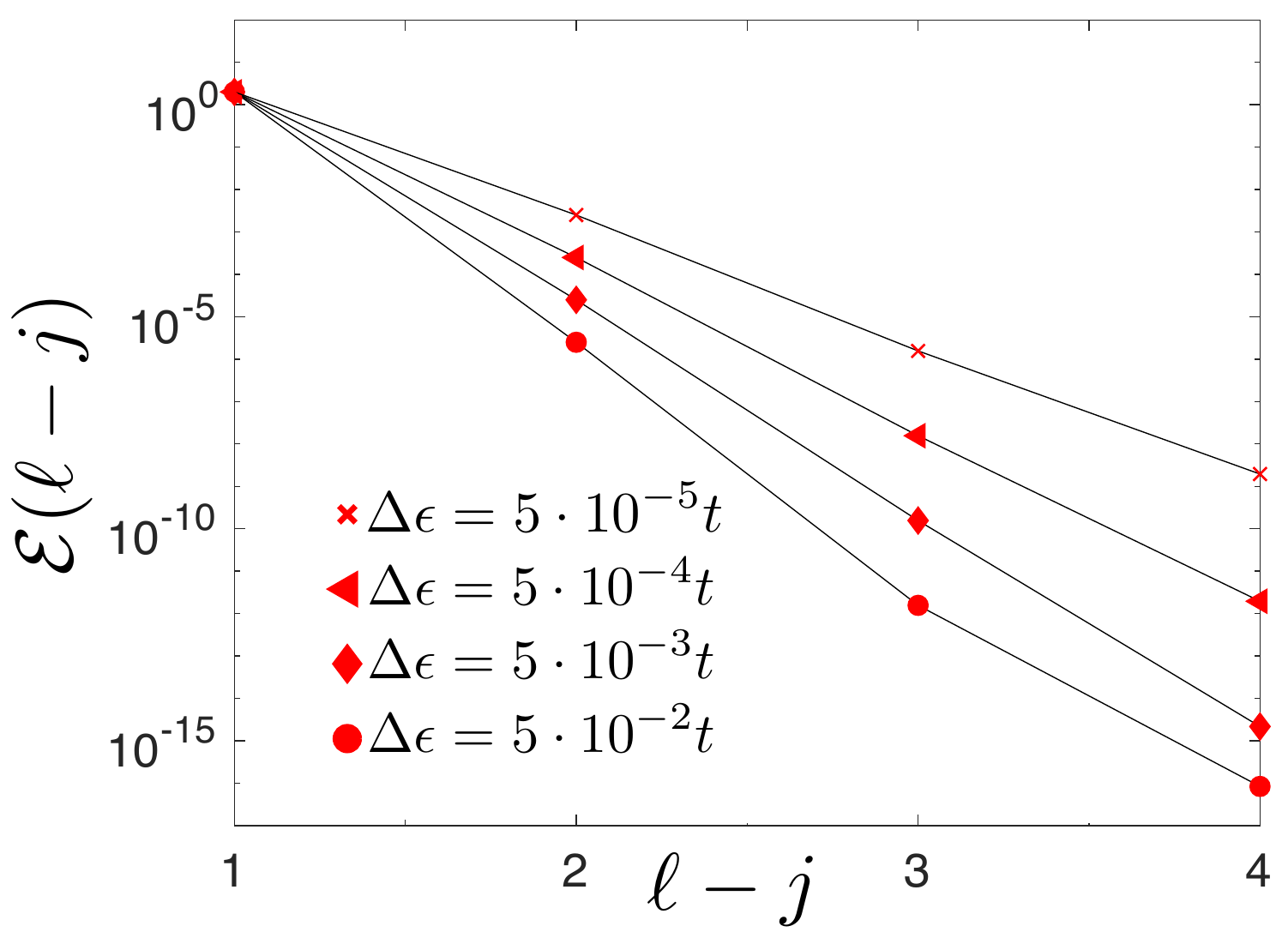}
  \caption{The effective coupling $\mathcal{E} (\ell-j)$ with $J=t$ for different values of $\Delta \epsilon$; 
  {{for $j=\ell$ we get a constant contribution which can be neglected.}}}
  \label{fig:bos}
\end{figure}
In order to probe the existence of a hierarchy of fully gapped phases at fractional fillings, 
we conveniently introduce the real-space fermionic operators 
$
  \hat d_{j,\eta} =\sqrt{L}^{-1} \sum_k e^{ikj} \hat d_{k,\eta}
$
built up from the momentum-space operators $\hat d_{k,\eta}$ defined in Eq.~(\ref{unitary}). Then we
remap the original fermionic operators $\hat c_{j,\sigma}$ onto the new ones $\hat d_{j,\eta}$ as 
\begin{equation}
\cases{
\hat c_{j,\uparrow} = \sum_\ell \left[ F_c(j-\ell) \hat d_{\ell, -1} - F_s(j-\ell) \hat d_{\ell, +1} \right] \\
\hat c_{j,\downarrow} = \sum_\ell \left[ F_s(j-\ell) \hat d_{\ell, -1} + F_c(j-\ell) \hat d_{\ell, +1}\right]}
\label{mapping}
\end{equation}
where
\begin{equation}
\cases{
F_s(j-\ell)= \frac{1}{L} \sum_k e^{ik(j-\ell)} {e^{-i \varphi_k}} \sin \frac{\theta_k}{2}  \\
F_c(j-\ell)= \frac{1}{L} \sum_k e^{ik(j-\ell)} {e^{-i \varphi_k}} \cos \frac{\theta_k}{2} \; }
\label{FCFS}
\end{equation}
{{are the Wannier functions of the tight-binding model}}; in the following, we assume $\varphi_k=0$. For this choice of $\varphi_k$, the Wannier functions $F_c(j-\ell)$ and $F_s(j-\ell)$ can be calculated exactly when $\Delta \epsilon=0$, as shown 
in~\ref{FcFs}. When $\Delta \epsilon>0$, the functions $F_c(j-\ell)$ and $F_s(j-\ell)$ can be calculated numerically. 
However, we have verified that the functions $F_c(j-\ell)$ and $F_s(j-\ell)$ exhibit a weak dependence on $\Delta \epsilon$ and their expressions calculated for $\Delta \epsilon=0$ are a good approximation as long as  $\Delta \epsilon \ll t$.

To simplify the problem, we project on the lowest band by assuming that the interaction terms { $U$ and $V_p$} are much smaller than the band gap, i.e. $\approx 4J$ when $J=t$ and $\Delta \epsilon=0$.  Since we are dealing with low fillings anyway, it is reasonable to suppose that only the lower band is significantly populated (from now on, we will thus omit the index $-1$). In order to check the self-consistency of our predictions, numerical simulations will nonetheless be performed with the full description of the system. 
Under these assumptions, $\hat H_0$ becomes 
\begin{equation}
  \hat H_0= \sum_{j, \ell} \Big[ \mathcal{E} (\ell-j)  \, \hat d_{j}^\dagger \hat d_{\ell} +\mathrm{H.c.} \Big]
  \label{H_nonlocal}
\end{equation}
with $\mathcal{E} (\ell-j) = \frac1L \sum_k E(k) \, e^{ik (\ell -j)}$.
Of course, the Hamiltonian~(\ref{H_nonlocal}) is highly non-local, since all sites are coupled together by long-range terms. 
However, as shown in Fig.~\ref{fig:bos}, the coefficient $\mathcal{E} (\ell-j)$ decays exponentially with $\ell-j$ and the lower band can be approximated by truncating to nearest-neighbor terms:
\begin{equation}
  \hat H_0 \approx - \mathcal{E}_{1} \sum_{j} \Big( \hat d_{j+1}^\dagger \hat d_{j}+ \mathrm{H.c.} \Big) , \label{HH0}
\end{equation}
where we have defined $\mathcal{E}_1 \equiv -\mathcal{E} (1)$ and neglected an inessential chemical potential.
For $J=t$, it turns out that $\mathcal{E}_1 = \Delta \epsilon/2$. We stress here that, a truncation up to nearest-neighbor terms only breaks the symmetries of the original model and the new Hamiltonian $\hat H_0$ is not topological. Nevertheless this approach is useful to show the appearance of a hierarchy of fully gapped phases. 
Their topological properties will be discussed in the following (see below). 

Let us  focus on the interaction terms.
By means of the mapping~(\ref{mapping}) and considering the dominant contributions, it is possible to approximate $\hat c_{j,\uparrow} \approx F_c(1) \hat d_{j+1} + F^*_c(1) \hat d_{j-1}$
and $\hat c_{j,\downarrow} \approx F_s(0) \hat d_{j}$. Then, the Hubbard interaction term $U \hat n_{j,\uparrow} \hat n_{j,\downarrow}$ is mapped onto a nearest-neighbor density-density interaction term of the form $\hat n_j \hat n_{j+1}$ with $\hat n_j=\hat d^\dagger_j \hat d_j$, plus additional contributions (see below). Similarly, the density-density terms 
$V_p\hat n_{j} \hat n_{j+p}$ in the original model will be mapped onto density-density terms of the form $\hat n_{j} \hat n_{j+p+1}$. 
These density-density interaction terms lead to a {{hierarchy of gapped phases}} supporting density-wave states at rational filling fractions --- the well-known Devil's staircase~\cite{Hubbard_78, Pokrovsky_78, Giamarchi_04}, which we now discuss in the context of our model. 
{{In the next paragraph~\ref{filling12} we address in detail the filling $\nu=1/2$, then we generalize our results to lower fillings and we explicitly consider the filling $\nu=1/3$ in par.~\ref{sec:DevilStair}.}}

\subsection{Topological density-wave at $\nu=1/2$: analytical and numerical characterization} 
\label{filling12}

\begin{figure}
  \includegraphics[width=0.95\columnwidth]{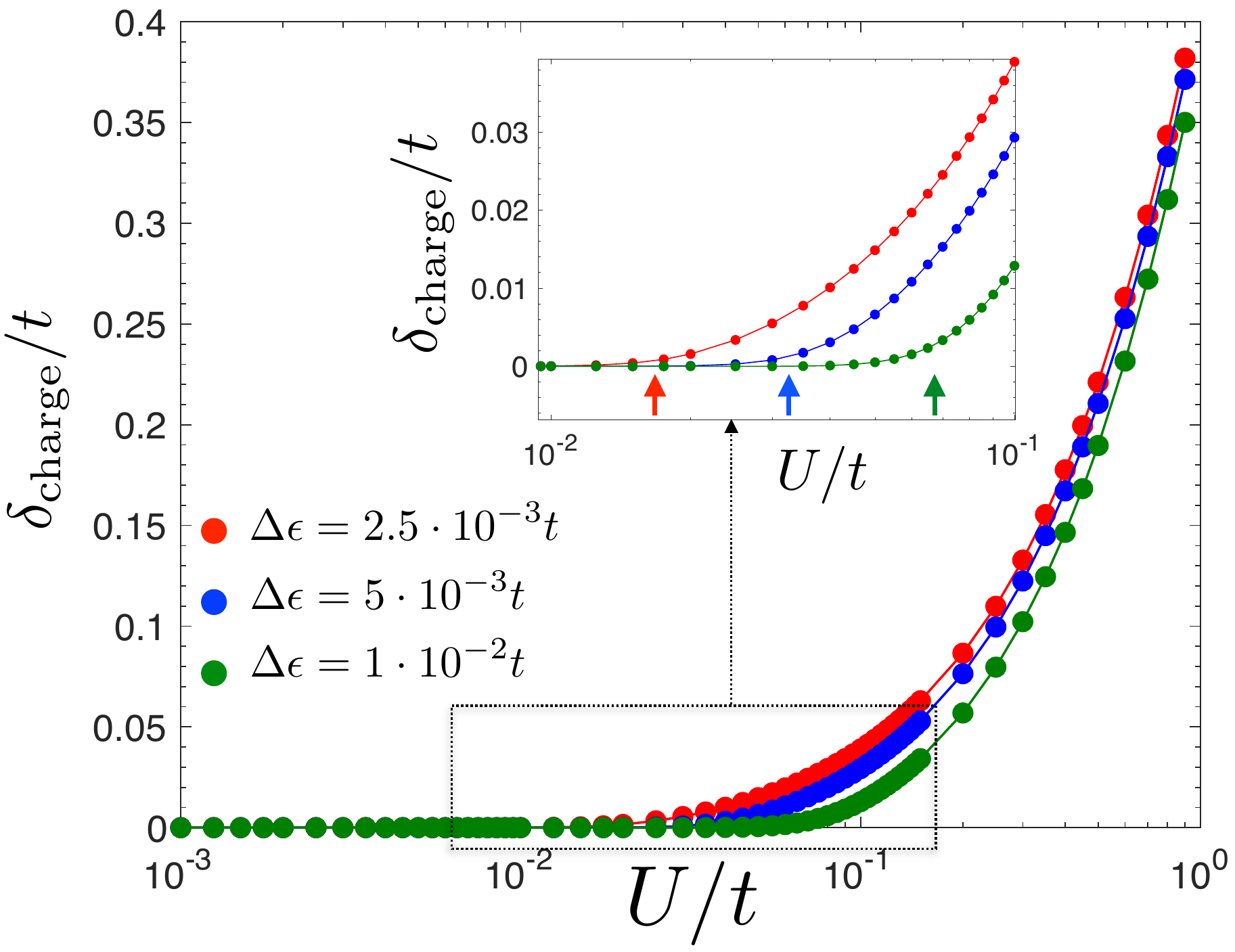}
  \caption{Charge gap $\delta_{\rm charge}$ at filling $\nu=1/2$ as a function of $U/t$, for different values of $\Delta \epsilon$,
    and assuming PBC. The gap $\delta_{\rm charge}$ is obtained by performing a finite-size scaling~\cite{finitesize}
    of numerical exact diagonalization data for $L=4, \ldots, 14$, with $\delta_{\rm charge}(L)=\delta_{\rm charge}+b/L+c/L^2$. 
    Inset: magnification of the region inside the box; arrows indicate the critical interaction $U_c$, see Eq.~(\ref{anal}).}
  \label{fig:gap}
\end{figure}

In this paragraph we focus on a fractional topological phase at filling $\nu=1/2$ whose appearance can be discussed in a transparent way, both analytically (by means of a mean-field approach) and numerically (using exact diagonalization and DMRG).
Firstly we estimate the critical interaction which stabilizes a gapped phase. To this aim we rewrite the interaction term $\hat n_{j,\downarrow} \hat n_{j,\uparrow}$ 
using the mapping~(\ref{mapping}) approximated at the first non-trivial order, i.e.
$\hat c_{j,\uparrow} \approx F_c(1) \hat d_{j+1} + F^*_c(1) \hat d_{j-1}$
and $\hat c_{j,\downarrow} \approx F_s(0) \hat d_{j}$
and projecting on the lowest band.
Omitting terms which vanish because of the Pauli principle, we obtain
\begin{equation}
  \hat H_{U} \approx \tilde U \sum_j \Big[ \hat n_j \hat n_{j+1} - \frac 12\big( \hat n_j \hat d^\dagger_{j+1} \hat d_{j-1} +\mathrm{H.c.} \big) \Big],
  \label{HUU}
\end{equation} 
with $\tilde U=2U |F_c(0)|^2 |F_s(1)|^2$. 
Within a mean-field approach, the correlated hopping term $ \hat n_j \hat d^\dagger_{j+1} \hat d_{j-1}$ can be neglected (see~\ref{app_mf} for details), and the effective Hamiltonian given by Eq.~(\ref{HH0}) plus the density-density interaction term of Eq.~(\ref{HUU}) is equivalent to a spin-$1/2$ XXZ model, which can be exactly solved~\cite{Takahashi}.
The critical interaction $U_c$ stabilizing an antiferromagnetic gapped phase is 
\begin{equation}
U_c \sim \frac{\Delta \epsilon}{2 |F_c(0)|^2 |F_s(1)|^2} ,
\label{anal}
\end{equation} 
where the functions $F_c$ and $F_s$ were defined in Eq.~(\ref{FCFS}); in the case of filling $\nu=1/2$, a gapped phase can be stabilized by the Hubbard interaction only, for this reason, in the following, longer range interaction terms  $V_p$ are set to zero.

To substantiate our analytic predictions, assuming PBC conditions, we have numerically computed by means of a Lanczos-based exact diagonalization approach the charge gap of the interacting Hamiltonian~(\ref{HBDI})
\begin{equation}
  \delta_{\rm charge} = E_1(N) - \frac 12 \big[ E_1(N+1)+E_1(N-1) \big] ,
\end{equation}
where $E_\alpha(N)$ is the energy of the $\alpha$-th state with $N$ particles ($E_1$ being the ground-state energy),
and here we set $N=L/2$.
Figure~\ref{fig:gap} displays $\delta_{\rm charge}$ as a function of the interaction parameter $U/t$, for different values of the imbalance term $\Delta \epsilon$. We observe a good agreement of the analytic prediction~(\ref{anal}) for the critical interaction $U_c$, with the point at which the charge gap closes.

Likewise, the charge neutral gap 
\begin{equation}
\delta_{{\rm spin}, \alpha}=E_\alpha(N)-E_1(N)
\label{neutral_gap}
\end{equation}
can be obtained following a similar procedure. In particular, since the ground state always exhibits a two-fold degeneracy (see below), we consider $\alpha=3$.
The behavior of the spin gap $\delta_{{\rm spin}, 3}$ as a function of the interaction parameter $U/t$  is qualitatively analogous to the one of the charge gap (data not shown).

\subsubsection{Ground-state degeneracy}

\begin{figure}
  \includegraphics[width=0.95\columnwidth]{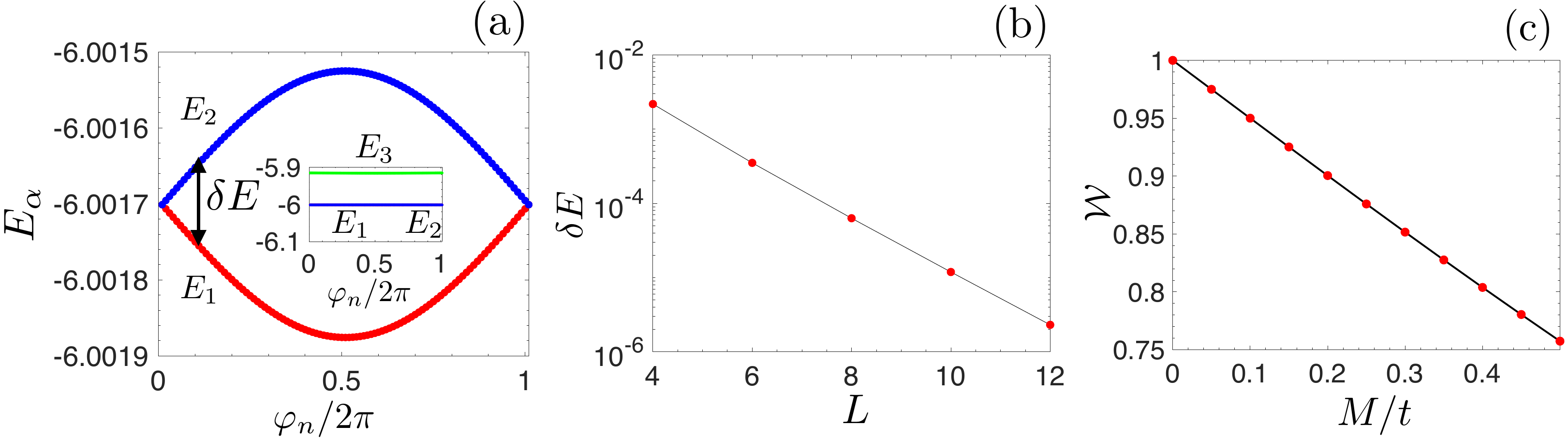}
  \caption{(a) Ground-state energies $E_1$ and $E_2$ as a function of the twisting angle $\varphi_n$.
    The inset displays the same plot, for the first three levels, on an enlarged energy scale energies.
    As expected, the gap between $E_1$ and $E_2$ and the first excited state $E_3$ is preserved when $\varphi_n \neq 0$. 
    (b) The difference $\delta E \equiv E_2-E_1$ scales exponentially with the chain length $L$ (here $\varphi_n=\pi$).
    (c) The Wilczek-Zee phase as a function of the symmetry-breaking Hamiltonian term $M$ of Eq.~(\ref{HSB}). As expected, for $M=0$, the Wilczek-Zee phase is quantized to one. 
    Data have been obtained through exact diagonalization, with $U=0.4t$, $\Delta \epsilon = 10^{-2}t$, and $N_\varphi=100$.
    Panel (a) is for $L=6$, while panels (c-d) are for $L=8$.
   }
  \label{fig:berry}
\end{figure}

Before addressing the topological properties of the fractional phase, it is worth investigating  
the spectrum $\{ E_\alpha(N)\}$ with $\alpha=1,\,2,\, \dots$ and $N=L/2$, 
for both PBC and open boundary conditions (OBC). Our numerics evidences how the ground-state degeneracy does indeed depend on the choice
of the boundary conditions: while for PBC it is doubly degenerate, see Fig.~\ref{fig:berry}(a), for OBC it is non-degenerate~\cite{Pollmann10,Stoudenmire_11,Kraus_13}. 

To explain the reason of this anomalous degeneracy, we perform the unitary transformation $\hat \mathcal{ U}_j \, \hat c_{j,\sigma} \, \hat \mathcal{ U}_j^{-1} = e^{i \frac{\pi}{2}j} \hat c_{j,\sigma}$ and introduce the ``cage operators''
\begin{equation}
  \hat \gamma^\dagger_{j,\pm} = \frac12 \big(i \hat c^\dagger_{j,\uparrow} +\hat c^\dagger_{j,\downarrow} \big) \mp \frac12 \big(\hat c^\dagger_{j+1,\uparrow} + i \hat c^\dagger_{j+1,\downarrow} \big) ,
\end{equation}
which span four lattice sites (of shape $2\times2$). Then, in the simple case $t=J$, $\Delta \epsilon=0$,
the Hamiltonian~(\ref{HBDI}) can be rewritten as 
\begin{equation}
\hat H= - 2t \sum_{j} \left( \hat \gamma^\dagger_{j,-} \hat \gamma_{j,-}-\hat \gamma^\dagger_{j,+} \hat \gamma_{j,+} \right) + \hat H_{U} .
\end{equation}
If we now consider the regime where $U \ll t$ and get rid of the upper band,
we realize that $\hat H_{U}$ corresponds to a nearest-neighbor interaction term between the cages, i.e. $\hat n_{j,-} \hat n_{j+1,-}$, with $\hat n_{j,-}= \hat \gamma^\dagger_{j,-} \hat \gamma_{j,-}$
(see also Ref.~\cite{Junemann_17}). Then, since the ground state at filling  $\nu=1/2$ can be schematically represented via the occupation of local cages, we observe that PBC can effectively fit two of those states (where the cages start at odd or even sites, respectively). Conversely, OBC can only accommodate a single one (where the cages start at odd sites). This interpretation also explains the robustness of the ground-state degeneracy when the boundary conditions are twisted in a closed chain, as the rigid cage structure is not sensitive to such a twist --- see next subsection. 
This behavior is akin to the robustness of ground-state degeneracy in true topologically ordered states (see also Ref.~\cite{Budich_13}).

\subsubsection{Wilczek-Zee phase}
{{As discussed in the previous paragraph, when PBC or twisted boundary conditions are assumed the ground state at filling $\nu=1/2$ is gapped and two-fold degenerate. For this reason}} the correct topological invariant which has to be used to reveal its topological properties is the Wilczek-Zee phase~\cite{Wilczek_84,Chruscinski2004,Niu_85}
\begin{equation}
  \mathcal{W}= \frac{i}{\pi} \int_0^{2\pi}  d\varphi \; \mathrm{Tr} \big[\langle \Psi_\alpha(\varphi)|\partial_\varphi|\Psi_\beta(\varphi)\rangle \big] ,
  \label{WZ}
\end{equation}
where $\{ |\Psi_\alpha(\varphi)\rangle \}$ are the different degenerate many-body ground states labeled by the index $\alpha=1,\,\dots,D$, here with $D=2$, while $\varphi$ is the twisting angle; 
{{$\alpha$ and $\beta$ are the indices over which the trace is performed}}.  Twisted boundary conditions along the physical dimension can be implemented by taking $t \rightarrow t \exp(i \varphi / L)$
and $J \rightarrow J \exp(i \varphi / L)$.
The quantity $\mathcal{W}$ in Eq.~(\ref{WZ}) can be numerically computed through the procedure of Ref.~\cite{Resta_94}:
one first discretizes the angle $\varphi \in [0, 2\pi]$ in $N_\varphi$ steps of $\delta \varphi=2\pi/N_\varphi$,
each corresponding to the value $\varphi_n$ ($n=0,\ldots,N-1$).
Then, after solving the Schr\"odinger equation $\hat H(\varphi_n)|\Psi_{n,\alpha} \rangle = E_\alpha(\varphi_n) |\Psi_{n,\alpha} \rangle$
at the $n$-th step, the obtained many-body ground states $|\Psi_{n,\alpha} \rangle$ can be used to build up the Berry connection
\begin{equation}
  A_n = \mathrm{Im} \log \mathrm{det}[ D^{\alpha,\beta}_n] \quad \mbox{with} \;
  D^{\alpha,\beta}_n= \langle \Psi_{n,\alpha}|\Psi_{n+1,\beta}\rangle.
\end{equation}
The WZ phase is defined by $\mathcal{W}=\sum_{n=0}^{N-1} A_n$.

First of all, in Fig.~\ref{fig:berry}(a) we plot the ground-state energies $E_1(\varphi_n)$ and $E_2(\varphi_n)$ as a function of the discretized twisting angle $\varphi_n$ and observe that the exact degeneracy at $\varphi_n=0$ is only apparently removed when $\varphi_n \neq 0$. Indeed, as shown in Fig.~\ref{fig:berry}(b), the difference $\delta E(\varphi_n=\pi)=E_2(\pi)-E_1(\pi)$ scales exponentially with the system size $L$. As expected, 
the inset of Fig.~\ref{fig:berry}(a) highlights that the neutral gap~(\ref{neutral_gap}) does not close when twisted boundary conditions are used, as it is essentially insensitive to boundary conditions. 
Figure~\ref{fig:berry}(c) demonstrates that, in the presence of a chiral symmetry-breaking Hamiltonian term 
\begin{equation}
\hat H_{\rm SB}=iM\sum_j \big( \hat c^\dagger_{j,\uparrow} \hat c_{j,\downarrow} -\mathrm{H.c.} \big) \,,
\label{HSB}
\end{equation}
i.e. $\hat \mathcal{ S} \, \hat H_{\rm SB} \, \hat \mathcal{ S}^{-1} \neq \hat H_{\rm SB}$,
the WZ phase is not quantized anymore, thus signaling that the fractional gapped phase is protected by the same symmetry of the integer case. {{ On the contrary, for $M=0$, 
the WZ phase is strictly quantized to one independently of the value of $\Delta \epsilon$ (as long as the interaction term $U$ is sufficiently strong to stabilize a gapped phase). }}

\subsubsection{Entanglement spectrum}

\begin{figure}
  \includegraphics[width=0.95\columnwidth]{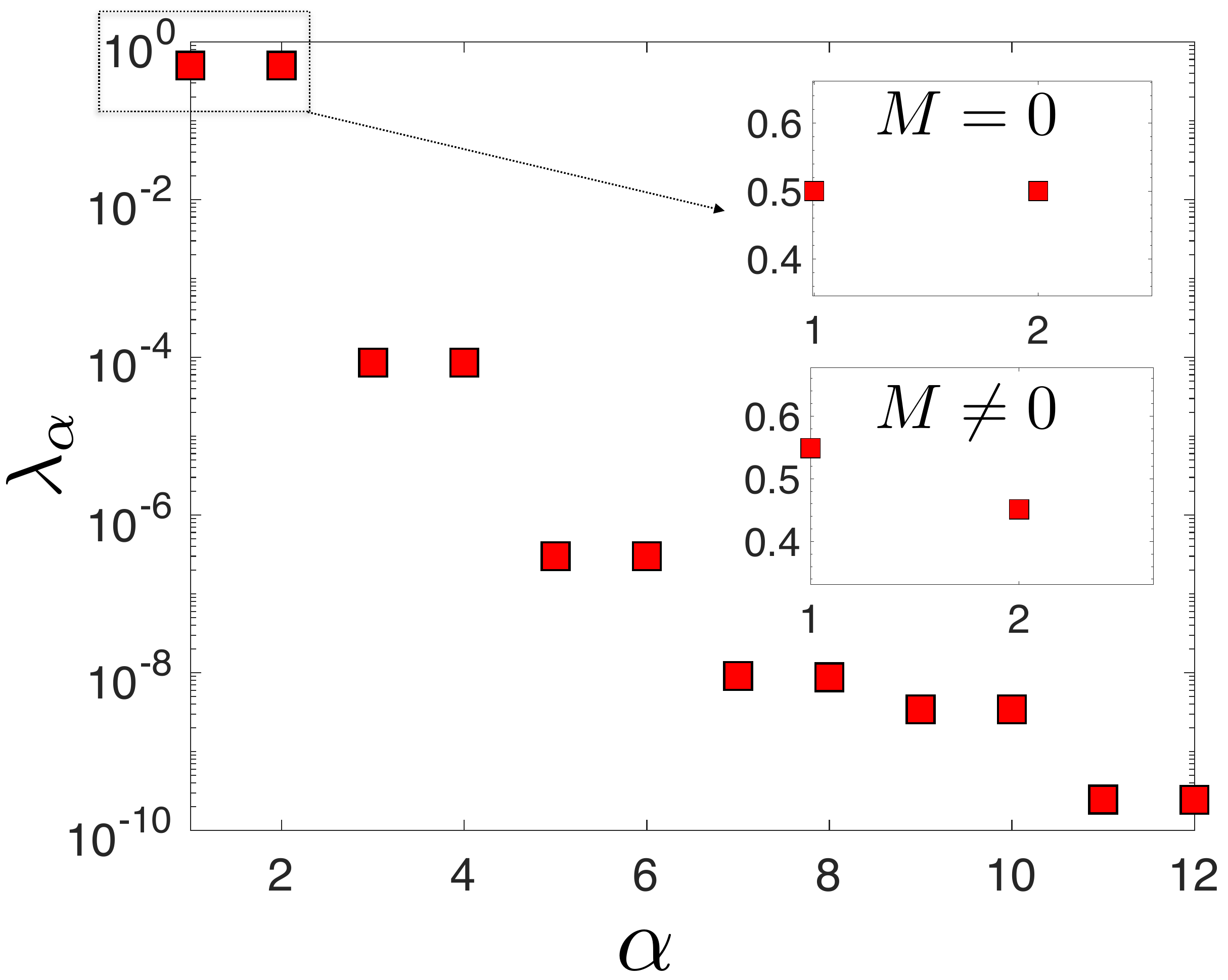}
  \caption{Entanglement spectrum for a chain of $L=100$ sites, $U=t$ and $\Delta \epsilon=2 \cdot 10^{-2}t$; 
    in the present case $\ell=L/2$. Upper inset: magnification of the two largest eigenvalues.
    Lower inset: the two largest eigenvalues in the presence of the chiral symmetry-breaking term~(\ref{HSB}), with $M=0.2\,t$.}
  \label{figure:ent}
\end{figure}

To substantiate the topological nature of the gapped phase discussed so far, we have investigated the entanglement spectrum {{ of the Hamiltonian~(\ref{HBDI}) by means of DMRG simulations.  }}
Such quantity corresponds to the set of the eigenvalues $\{ \lambda_\alpha \}$ of the reduced density matrix $\hat \rho_\ell=\mathrm{Tr}_{\overline{\ell}} \big[ |\Psi\rangle \langle \Psi| \big]$ obtained from the system's ground state $|\Psi\rangle$.
Here we consider a subsystem containing $\ell < L$ adjacent sites, and call $\overline{\ell}$ its complement;
{{we note that the degeneracy of the entanglement spectrum is not altered when other values of $\ell$ are considered. }}
It is well known~\cite{Pollmann10, Fidkowski10} that there exists a connection between the topological vs.~trivial nature of $|\Psi\rangle$ and the degeneracy of the eigenvalues of $\hat \rho_\ell$. A topological phase corresponds to a degenerate entanglement spectrum: this is indeed what we observe in Fig.~\ref{figure:ent}, where we plot the first twelve eigenvalues of the entanglement spectrum for a chain of $L=100$ sites, $U=t$ and $\Delta \epsilon=2 \cdot 10^{-2}t$. The upper inset is a magnification of the two largest eigenvalues, whose degeneracy is removed in the presence of the symmetry breaking Hamiltonian term~(\ref{HSB}) --- see the lower inset.

\subsubsection{Unconventional edge physics at $\nu=1/2$}

Topological phases are typically characterized by the presence of zero-energy modes, when OBC along the physical dimension are assumed.
A necessary but non sufficient condition for their presence is a vanishing (resp. non-vanishing) single-particle charge gap at filling $\nu=1/2$ with OBC (resp. PBC).
Here, despite the topological nature of the model, zero-energy modes do not appear, since the single-particle charge gap
remains finite even with OBC, and exhibits a behavior qualitatively analogous to the PBC case --- see Fig.~\ref{fig:gap}. 

Although zero-energy modes are absent, the topological nature of the model manifests itself in an unconventional edge physics which can be revealed through the quantity 
\begin{equation}
  \delta n_j = \langle L/2+1|\hat n_j |L/2+1\rangle - \langle L/2|\hat n_j |L/2\rangle ,
\end{equation}
measuring the difference between the expectation value of the density operator onto the state $|L/2\rangle$ corresponding to filling $\nu=1/2$,
and its expectation value onto the state $|L/2+1\rangle$ corresponding to filling $\nu=1/2+1/L$.

\begin{figure}
  \includegraphics[width=0.95\columnwidth]{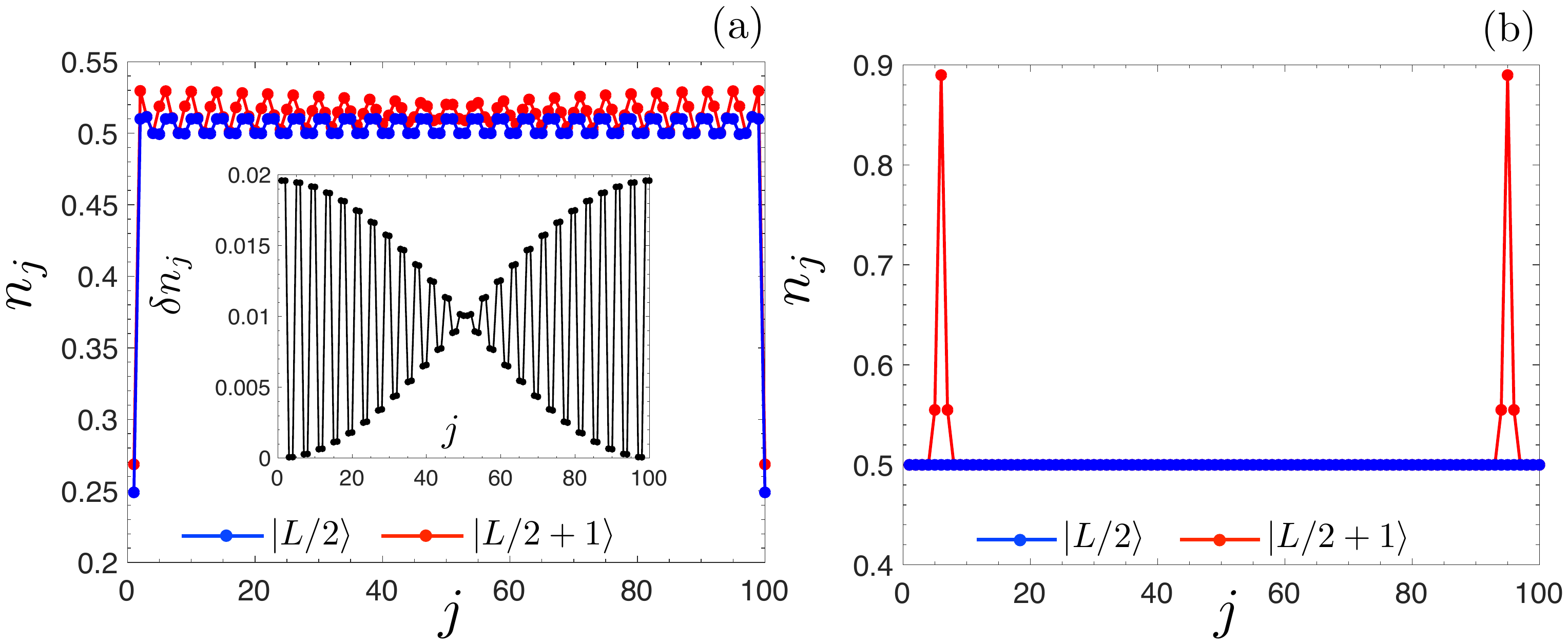}
  \caption{Expectation value of the density operator $\hat n_j$ at $\nu=1/2$ (blue) and $\nu=1/2+1/L$ (red),
    for $U=0$ [panel (a), exact diagonalization] and for $U=t$ [panel (b), DMRG].
    The inset shows the quantity $\delta n_j$. 
    Here $J=0.99t$, $\Delta \epsilon=0$, and $L=100$.}
  \label{figure:density}
\end{figure}

\begin{figure}
  \includegraphics[width=0.95\columnwidth]{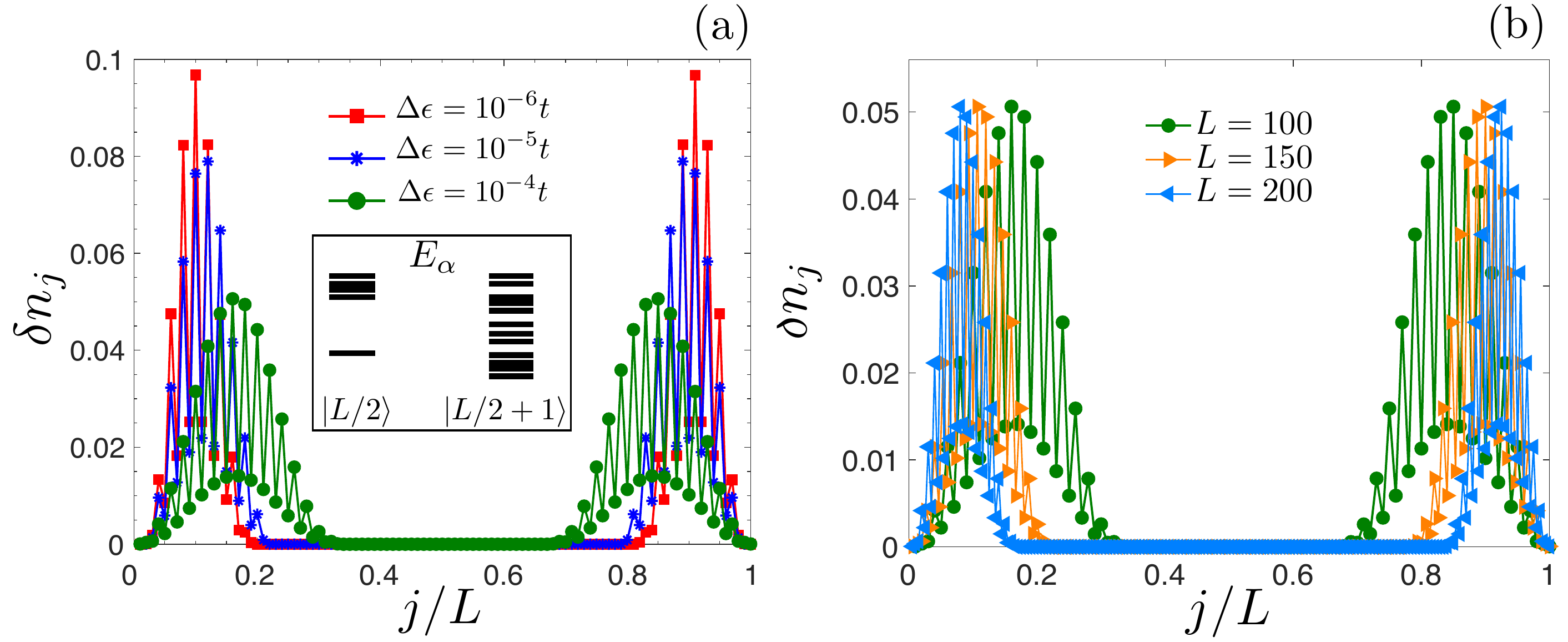}
  \caption{The quantity $\delta n_j$ for different values of the imbalance $\Delta \epsilon$ and fixed $L=100$ [panel (a)],
    and for different sizes $L$ and fixed $\Delta \epsilon=10^{-4}t$ [panel (b)].
    Here $J=0.99t$, $U=t$. Data have been obtained by means of DMRG.}
  \label{figure:de}
\end{figure}

We start investigating the case where the spin imbalance $\Delta \epsilon$ vanishes.
In Fig.~\ref{figure:density} we plot the two density profiles both in the non-interacting case where the phase is gapless, and in the interacting case, for a sufficiently large interaction term $U$ which stabilizes the topological gap. 
In the non-interacting case [panel (a)], where the quantity $\delta n_j$ describes the wave-function of the added particle, no edge physics is observed, since $\delta n_j$ is delocalized over the entire chain. On the contrary, in the interacting case [panel (b)], $\delta n_j$ displays two sharp peaks close to the edges of the system. 
When a small imbalance term $\Delta \epsilon$ is considered, the behavior of $\delta n_j$ changes drastically.
As shown in Fig.~\ref{figure:de}, we still observe some edge physics, but the quantity 
$\delta n_j $ is now spread over a number of sites that increases with growing $\Delta\epsilon$.

We now give an intuitive picture of this spreading effect. To this aim we consider the inset of Fig.~\ref{figure:de}(a) where a qualitative picture of the spectrum $\{E_1, E_2, \dots \}$ of the interacting Hamiltonian is shown
at  $\nu=1/2$  and at  $\nu=1/2+1/L$, with $\Delta \epsilon=0$ and OBC. 
In the first case $|L/2 \rangle$, there is a finite gap between the unique ground state and the first excited state.
When the local imbalance term $\Delta \epsilon$ is added, the ground state is unmodified as long as $\Delta \epsilon$ is small with respect to the gap.
In the second case $|L/2 +1\rangle$, the ground state is not protected by a finite energy difference.
For this reason, in the presence of the imbalance term $\Delta \epsilon$, it is expected to be a quantum superposition of the ground state $|L/2 +1\rangle$ at $\Delta \epsilon=0$ plus pieces coming from the excited states which carry bulk contributions, and originate the spreading of the quantity $\delta n_j$ shown in Fig.~\ref{figure:de}.  
However, the spreading of $\delta n_j$ in the bulk becomes negligible in thermodynamic limit, as shown in Fig.~\ref{figure:de}(b) for different sizes of the chain.

\subsection{Devil's staircase from bosonization}
\label{sec:DevilStair}

So far we have considered the fractional topological phase at filling $\nu=1/2$. 
The appearance of a hierarchy of topological gapped phases at lower fillings can be explained in terms of a bosonization approach~\cite{Giamarchi_04}. 
To this aim, we consider the continuum limit of the fermionic operators $\hat d_j$, definined by $\hat d_j \equiv \hat d(x)/\sqrt{a}$
and $\hat d_{j+1} \equiv \hat d(x+a)/\sqrt{a}$, with $a$ being a generic cut-off length (in the following, $a=1$).
This operator can be expressed in terms of the bosonic fields $\hat \phi(x)$ and $\hat \theta(x)$
satisfying $[\hat \phi(x), \partial_{x'} \hat \theta(x')]=i\pi \delta (x-x')$ as
\begin{equation}
\hat d(x)= \sum_p A_p \, e^{-i(2p+1) \left[ k_F x-\hat \phi(x) \right] }\, e^{i \hat \theta (x)},
\end{equation}
with $k_F = \pi N/L$ being the Fermi momentum. Moreover, the density operator $\hat \rho(x)= \hat d^\dagger(x) \hat d(x)$ is given by
\begin{equation}
\hat \rho(x)= -\frac{\partial_x \hat \phi(x)}{\pi} + \sum_{p\neq 0}  B_p \, e^{2ip \left[k_F x-\hat \phi(x)\right]};
\end{equation}
here $A_p$ and $B_p$ are non-universal coefficients which depend on the cut-off length of the theory. Within bosonization, the Hamiltonian~(\ref{HH0}) plus the density-density interaction terms can be recast into a quadratic form 
\begin{equation}
\hat H_{\rm bos}=  \frac{u}{2\pi} \int dx \left[ \frac{1}{K}(\partial_x \hat \phi)^2 +  K(\partial_x \hat \theta)^2 \right]  
\end{equation}
describing a critical gapless theory, plus a sum of sine-Gordon terms
\begin{equation}
\hat H_{\rm sg} =  \sum_{p>2} M_p \int dx\; \cos \! \big[ 2p(\hat \phi(x) + k_Fx) \big] ,
\label{SG}
\end{equation}
where $u$ is an effective Fermi velocity, $K$ is related to the strength of the interaction terms, while the coefficients $M_p$ represent the amplitudes of the sine-Gordon terms. 
An exact mapping of the quantities $u$, $K$, and $M_p$ onto the microscopic parameters is beyond the scope of the present discussion and is generally challenging, due to the complex non-local character of the effective interactions. Furthermore, as shown in~\ref{app_bos}, at filling $\nu=1/2$, all interaction terms which cannot be recast into a density-density form and which have been so far neglected,  lead to a renormalization of the coefficients $u$, $K$ and $M_p$ only. 

\begin{figure}
  \includegraphics[width=0.95\columnwidth]{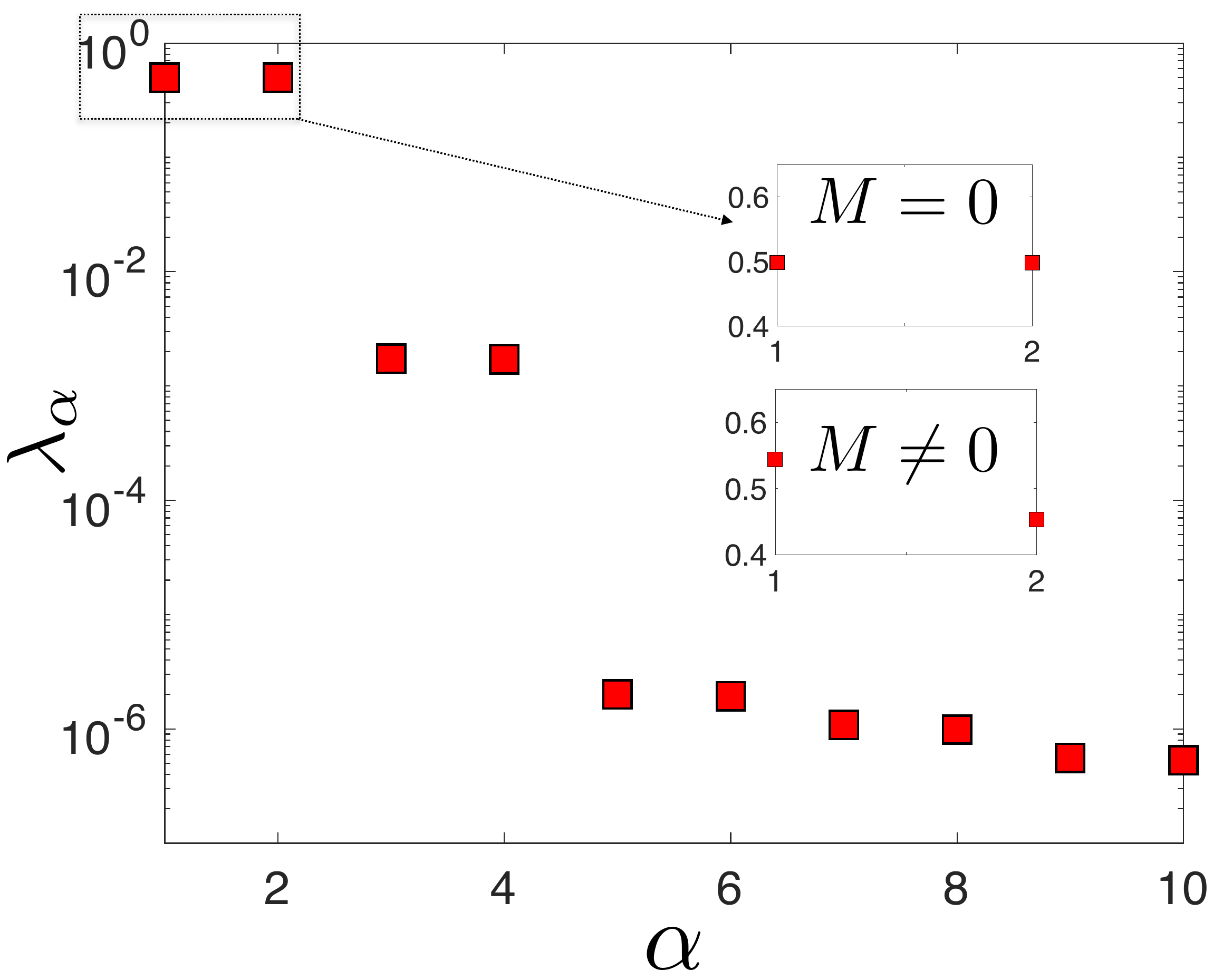}
  \caption{Entanglement spectrum for a chain of $L=60$ sites, $U=0$, $V_1=t$ and $\Delta \epsilon=8 \cdot 10^{-2}t$ at filling $\nu=1/3$. 
    Upper inset: magnification of the two largest eigenvalues.
    Lower inset: the two largest eigenvalues in the presence of a chiral symmetry-breaking term~(\ref{HSB}), with $M=0.2\,t$.}
  \label{figure:ent13}
\end{figure}

Sine-Gordon terms~(\ref{SG}) are responsible for the appearance of the gapped phases at fractional fillings. Indeed, when the space dependent $2p k_Fx$ term in the co-sinusoidal functions vanishes, i.e. $2p k_F \propto 2\pi$, they become relevant for $K<2/p^2$ and open a gap.
We stress that, within the present bosonization approach, we cannot say anything about the topological properties of these phases.
In the case of filling $\nu=N/L=1/2$, i.e. $k_F=\pi/2$, the most relevant sine-Gordon term is the one with $p=2$. 
All other phases at lower filling fractions $\nu=1/q$, with $q>2$, can be reached by considering sufficiently long-range density-density interaction terms, as in the conventional Devil's staircase scenario~\cite{Hubbard_78, Giamarchi_04, Dalmonte_10}. {{We now explicitly consider the Hamiltonian~(\ref{HBDI}) and discuss the fractional filling case $\nu=1/3$ for which the topological phase is stabilized by a nearest-neighbor interaction term in Eq.~(\ref{Hint}). In Fig. \ref{figure:ent13}, we plot the first ten eigenvalues of the entanglement spectrum for a chain of $L=60$ sites, $U=t$, $V_1=t$, and $\Delta \epsilon=8 \cdot 10^{-2}t$. As expected, the two highest eigenvalues are degenerate due to the topological nature of the ground state. Similarly to the case studied previously, we finally observe that their degeneracy is removed in the presence of the symmetry breaking Hamiltonian term~(\ref{HSB}), as shown in the lower inset, signaling that the fractional topological phase is protected by the same symmetry that protects the non-interacting topological phase at integer filling.
 }}

\section{Inversion symmetric topological phases at fractional fillings}
\label{inversion_results}

It is a natural question to inquire whether the mechanism for the stabilization of interaction-induced topological phases at fractional filling fraction is interwound with spatial symmetries (which play a key role in the establishment of conventional Devil's staircase structures). In this section, we discuss an interacting, crystalline topological insulator, where a fractional topological phase appears when considering interaction effects on the top of partly filled topological bands.

In particular, we consider a two-leg ladder which supports, in the non-interacting regime, a crystalline topological phase at filling $\nu=1$. 
Following the prescriptions given in Fig.~\ref{fig:model}(b), the resulting Hamiltonian reads~\cite{Hughes_11}:
\begin{equation}
\hat H_0 =  \sum_{\sigma, \, j} \left[ (\sigma t + \delta) \, \hat c^\dagger_{j+1,\sigma} \hat c_{j,\sigma} + 
iJ\, \hat c^\dagger_{j+1,\sigma} \hat c_{j,-\sigma} + \mathrm{H.c.} \right] + \Delta \epsilon \sum_{\sigma, \, j} \sigma \, \hat c^\dagger_{j,\sigma} \hat c_{j,\sigma} \,, 
\label{inversion_hamiltonian}
\end{equation}
with $\delta \neq 0$. This Hamiltonian, which is not endowed with a particle-hole symmetry nor a chiral symmetry, is characterized by the presence of edge states, by a quantized Zak phase, and by a doubly degenerate entanglement spectrum. Indeed the emerging topologcal phase at filling one is protected by a spatial inversion symmetry~\cite{Hughes_11} $\hat \mathcal{ I}$ which acts onto the fermionic operators $\hat c_{j,\sigma}$ as $\hat \mathcal{ I} \,\hat c_{j,\sigma}\,  \hat \mathcal{ I}^{-1} = I_{\sigma',\sigma} \hat c_{-j,\sigma'}$ such that $\hat \mathcal{ I} \, \hat H \, \hat \mathcal{ I}^{-1} = \hat H$ with $I=\sigma_z$.

In analogy with our previous discussion, we now show that an inversion symmetry protected topological phase at filling $\nu=1/2$ is stabilized by an on-site repulsive interaction term $\hat H_U$ as in Eq.~(\ref{Hint}).
In order to probe the emerging topological properties, we calculate the ground-state WZ phase~(\ref{WZ}) following the same procedure discussed for the BDI and AIII symmetry classes. 

\begin{figure}
  \includegraphics[width=0.95\columnwidth]{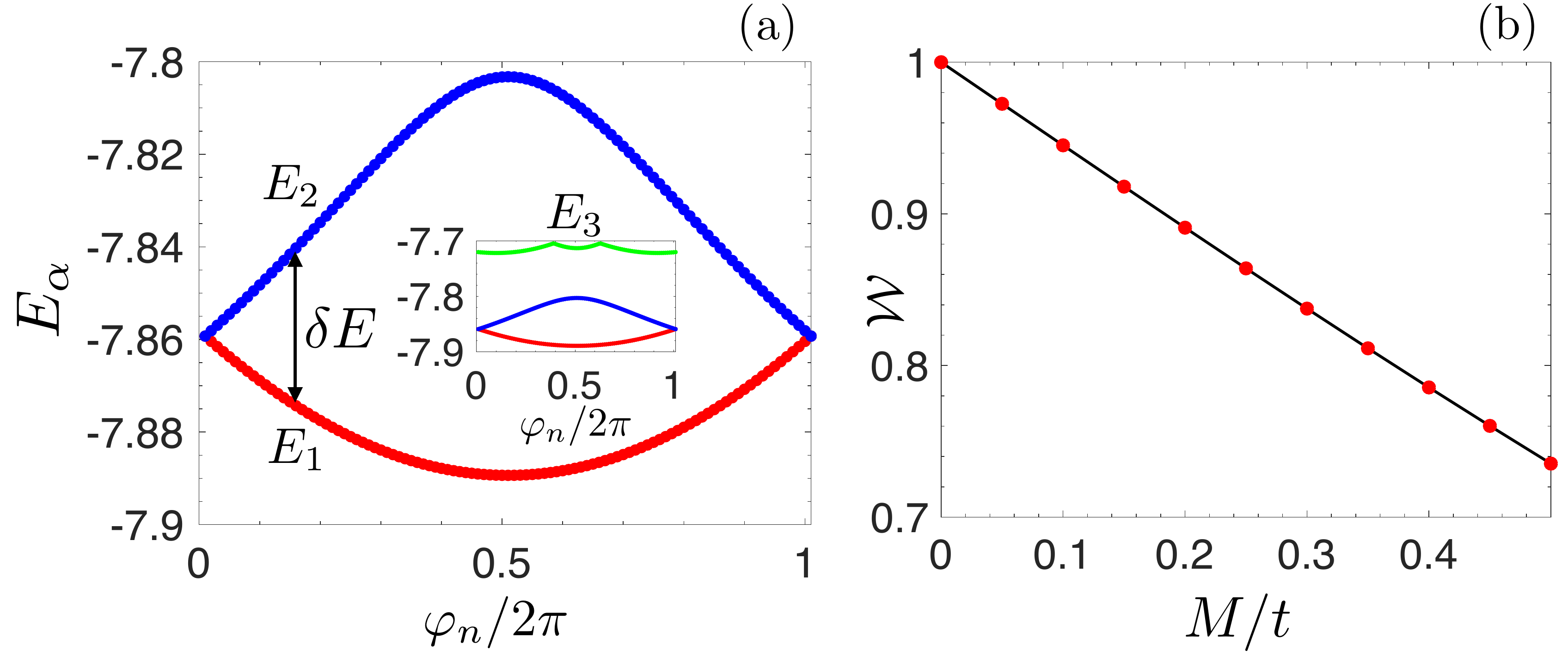}
  \caption{Same plot as in Fig.~\ref{fig:berry}(a) and~\ref{fig:berry}(d), but for the Hamiltonian
    of Eq.~(\ref{inversion_hamiltonian}) and its symmetry-breaking term $M$ of Eq.~(\ref{HSBinv}).
    Notice that, analogously to the case studied before, the gap between $E_1$ and $E_2$ and the first excited state $E_3$
    is preserved when $\varphi_n \neq 0$. Moreover, under the perturbation $M$, the WZ phase is not quantized to one anymore.
    Data have been obtained through exact diagonalization, with $L=8$, $\delta=0.1t$, $\Delta \epsilon=10^{-2}t$, $U=t$, $N_\varphi=100$.}
  \label{figure_inversion}
\end{figure}

In Fig.~\ref{figure_inversion}(a) we plot the energies $E_1(\varphi_n)$ and $E_2(\varphi_n)$ of the ground states as a function of the discretized twisting angle $\varphi_n$ for $\delta=0.1t$, $\Delta \epsilon=10^{-2}t$, and $U=t$. The inset shows that the first excited state $E_3(\varphi_n)$ is separated from the ground state by a finite gap. Similarly to the previously studied cases, the exact double degeneracy at $\varphi_n=0$ of $E_1$ and $E_2$ is only apparently removed when $\varphi_n \neq 0$. Indeed, the difference $\delta E(\varphi_n=\pi)=E_2(\pi)-E_1(\pi)$ scales exponentially with the system size (not shown). 

Then we introduce an Hamiltonian term 
\begin{equation}
  \hat H_{\rm SB}=M\sum_j \big( \hat c^\dagger_{j,\uparrow} \hat c_{j,\downarrow} +\mathrm{H.c.} \big) ,
  \label{HSBinv}
\end{equation}
which explicitly breaks the inversion symmetry, i.e. $\hat \mathcal{ S} \, \hat H_{\rm SB} \, \hat \mathcal{ S}^{-1} \neq \hat H_{\rm SB}$, and consequently the WZ phase is not quantized anymore. 

In Fig.~\ref{figure_inversion}(b) we plot the WZ phase as a function of the inversion symmetry breaking term $M$. As expected, for $M=0$ it is quantized and equal to one, while for $M\neq 0$ it is not quantized. This signals that the fractional gapped phase
is protected by same symmetry of the integer case, in complete analogy with what observed for the BDI/AIII cases.

\section{Conclusions}
\label{conclusions}
We have considered a 1D ladder with two internal spin states supporting a topological phase at integer filling and we have shown that, when the particle filling is reduced to a fractional value, repulsive interactions can stabilize a hierarchy of fully gapped density-wave phases with topological features.  

In particular we have focused on a specific example in the BDI class (unitarily equivalent to a model in the AIII symmetry class) of the Altland-Zirnbauer classification and on a crystalline topological insulator, i.e. a topological model protected by the spatial inversion symmetry. By means of a bosonization approach we have discussed the appearance of a gapped phase at filllings $\nu=1/q$ and, using exact numerical methods (DMRG simulations and exact diagonalization), we have verified our analytical predictions and we have also characterized the topological properties of the gapped phases at fillings  $\nu=1/2$ and $\nu=1/3$ by studying the topological quantum number (Wilczek-Zee phase) and the degeneracy of the entanglement spectrum. Considering the effects of perturbations, we have discussed how these fractional topological phases are protected by the same symmetry that protects the non-interacting topological phase at integer filling.

Most importantly, we have shown that these topological density waves do not follow the bulk-edge correspondence, in the sense that they exhibit modes at finite energy localized close to the edges of the system. Their presence has been diagnosed by studying the behavior of the density profile, when an extra particle is put in the system with respect to the filling $\nu=1/2$. Our results are immediately testable in cold atom experiments described by the setup in Ref.~\cite{Kang18,Kang18b}: while the single particle Hamiltonian has already been realized, a key requirement is to reach density regimes where an incompressible phase is stabilized in the center of the harmonic trap. 
Given that fractional phases appear already for quarter-filled band, we expect signal-to-noise not to constitute a problem. Since the incompressible phase in this regime has a gap of order $U$, this requires cooling in the tens of nanokelvin regime, which is within current experimental reach in these systems~\cite{Mancini15}.

We leave as an intriguing perspective the study of the appearance of these fractional topological phases in topological models belonging to the symmetry classes D, CII, and DIII of the AZc.

\ack
We thank the CINECA award under the ISCRA initiative for the availability of high performance computing resources and support. MD is supported by the ERC under grant number 758329 (AGEnTh). MD thanks ESI for hospitality during the Quantum Paths Programme. We acknowledge enlightening discussions with  Jan Budich, Alexander Nersesyan, and Lorenzo Pastori.


\appendix
 \section{Analytical calculation of  $F_c(j-\ell)$}
\label{FcFs}

In this Appendix we calculate analytically the functions 
\begin{equation}
\cases{
F_c(j-\ell)= \frac{1}{L} \sum_k e^{ik(j-\ell)} \cos \frac{\theta_k}{2} \\
F_s(j-\ell)= \frac{1}{L} \sum_k e^{ik(j-\ell)} \sin \frac{\theta_k}{2} }
\end{equation}
for $J=t$ and $\Delta \epsilon=0$. In this special case, $\cos \theta_k/2= \cos k/2$. Then, taking into account that
\begin{equation}
\sum_{k} e^{ikj} = \sum_{n=0}^{L-1} e^{\frac{2\pi i j}{L}n} \equiv \sum_{n=0}^{L-1} x^n = 
\cases{
L \;\;\; \mathrm{if} \;x=1\\
\frac{1-x^L}{1-x} \;\;\; \mathrm{if}\;  x\neq 1}
\end{equation}
with $x=e^{\frac{2\pi i j}{L}}$, we obtain
\begin{equation}
\cases{
F_s(j-\ell)=\frac{2}{L} \frac{ e^{\frac{2\pi i (j-\ell)}{L}} \sin \pi/L}{1+e^{\frac{4\pi i (j-\ell)}{L}}-2e^{\frac{2\pi i (j-\ell)}{L}} \cos \frac{\pi}{L}}\\
F_c(j-\ell)=\frac{2}{L} \frac{1-e^{\frac{2\pi i (j-\ell)}{L}} \cos \pi/L}{1+e^{\frac{4\pi i (j-\ell)}{L}}-2e^{\frac{2\pi i (j-\ell)}{L}} \cos \frac{\pi}{L}}\;.}
\end{equation}

\section{Bosonization at filling $\nu=1/2$}
\label{app_bos}

We discuss how the effective Hamiltonian 
\begin{eqnarray}
&\hat H \approx \hat H_0 + \hat H_U =-\frac{\Delta \epsilon}{2}\sum_{j} \left( \hat d_{j+1}^\dagger \hat d_{j,}  +\mathrm{H.c.} \right)+\nonumber\\
&+\tilde U \sum_j \Big[ \hat n_j \hat n_{j+1} - \frac 12\big( \hat n_j \hat d^\dagger_{j+1} \hat d_{j-1} +\mathrm{H.c.} \big) \Big],
\end{eqnarray} 
can be attacked by means of a bosonization approach. We introduce the continuum limit operators $\hat d(x)$ and $\hat d^\dagger(x)$ such that 
 $ \hat d_j \equiv \hat d(x)/\sqrt{a}$ and  $\hat d_{j+1} \equiv \hat d(x+a)/\sqrt{a}$, then the Hamiltonian $\hat H_0$ becomes
 \begin{equation}
\hat H_0 =  -\frac{\Delta \epsilon}{2} \int dx \left[   \hat d^\dagger(x+a) \hat d(x) + \mathrm{H.c.} \right]  \,;
\end{equation}
while the interaction term $\hat H_U \equiv \hat H_{U}^{(1)}+\hat H_{U}^{(2)} $ becomes
\begin{eqnarray}
&\hat H_{U}^{(1)} = \tilde Ua \int dx  \; \hat n(x) \hat n(x+a)  \\
&\hat H_{U}^{(2)} = -\frac 12 \tilde Ua \int dx   \left[ \hat n(x) \hat d^\dagger(x+a) \hat d(x-a) + \mathrm{H.c.} \right] 
\end{eqnarray}
here $\tilde U =2U F_c^2(0) |F_s(1)|^2$, while $a$ is the cut-off length of the theory.

To pursue a bosonization approach  we introduce the linearized (around the Fermi momentum $k_F$) fermionic operators $\hat d_R(x)$ and $\hat d_L(x)$ such that the original fermionic operator $\hat d(x)$ can be expanded as
$
\hat d(x) \approx e^{ik_Fx} \hat d_R(x) + e^{-ik_Fx} \hat d_L(x)
$
and rewritten as
$
\hat d_R(x)=(2\pi a)^{-1/2} e^{i[-\hat \phi(x)+\hat \theta(x)]}$ and  $\hat d_L(x)=(2\pi a)^{-1/2}  e^{i[\hat \phi(x)+\hat \theta(x)]} 
$
in terms of  the bosonic fields satisfying the usual commutation relation $[\hat \phi(x), \partial_x \hat \theta(x')] = i\pi \delta(x-x')$; the density operator $\hat n(x)= \hat d^\dagger(x) \hat d(x)$ is 
\begin{equation}
\hat n(x)=-\frac{1}{\pi} \partial_x \phi(x) +e^{-2ik_Fx} \hat d^\dagger_R(x) \hat d_L(x) + \mathrm{H.c.}  \,;
\label{density}
\end{equation}
moreover $\partial_x \hat \phi(x) = -\pi [\hat d^\dagger_R(x) \hat d_R(x) + \hat d^\dagger_L(x) \hat d_L(x)  ] $,
$\partial_x \hat \theta(x) = \pi [\hat d^\dagger_R(x) \hat d_R(x) - \hat d^\dagger_L(x) \hat d_L(x)  ]$.

Within the bosonization formalism, the non-interacting Hamiltonian $\hat H_0$ becomes
\begin{equation}
\hat H_0=  \frac{v_F}{2\pi} \int dx \left[ (\partial_x \hat \phi)^2 +  (\partial_x \hat \theta)^2 \right]
\end{equation}
where $v_F= \frac{a \Delta \epsilon}{2} \sin k_F a$. We recall here that $\nu=N/L=1/2$ and $k_F=\pi/(2a)$; then $e^{\pm 2i k_F a} = -1$. 
The bosonization of the Hamiltonian term $\hat H^{(1)}_{U}$ is quite standard, see e.g. Ref.~\cite{Giamarchi_04}, and leads to
 \begin{equation}
 \hat H^{(1)}_{U} = 2\tilde U a \left[ \frac{2}{\pi^2} (\partial_x \hat \phi)^2  - \frac{2}{(2\pi a)^2} \cos 4\hat \phi(x) \right]  \,.
 \end{equation}
 
The bosonization procedure of $\hat H^{(2)}_{U}$ is a bit more subtle. 
We preliminary consider the quantity $\hat d^\dagger(x+a) \hat d(x-a) + \mathrm{H.c.}$ that, up to a proper shift, can be rewritten as
$\hat d^\dagger(x+2a) \hat d(x) + \mathrm{H.c.}$ Then we expand it in terms of the right and left operators taking into account that 
$
 \hat d_r(x+2a) \approx \hat d_r(x) +2a \partial_x \hat d_r(x)
$,
 with $r=R,\, L$.  At the end of this procedure, we get four contributions $\hat h_{A,1}$, $\hat h_{A,2}$, $\hat h_{A,3}$, and $\hat h_B$ with
 \begin{eqnarray}
 & \hat h_{A,1} =-2\hat d_R^\dagger (x) \hat d_R(x) -2\hat d_L^\dagger (x) \hat d_L(x) \\
 & \hat h_{A,2} =-2a\hat d_R^\dagger (x) \partial_x \hat d_R(x) -2a \hat d_L^\dagger (x) \partial_x \hat d_L(x) \\
 & \hat h_{A,3} = -2a\partial_x \hat d_R^\dagger (x)  \hat d_R(x) -2a \partial_x \hat d_L^\dagger (x)  \hat d_L(x) \\
& \hat h_B = -e^{-2ik_Fx } \left[\hat d^\dagger_R(x) \hat d_L(x+2a) +\hat d^\dagger_R(x+2a) \hat d_L(x)\right] +\nonumber\\
&-e^{2ik_F x} \left[ \hat d^\dagger_L(x) \hat d_R(x+2a) + \hat d^\dagger_L(x+2a) \hat d_R(x) \right]\,.
 \end{eqnarray}
 Now, recalling the identity~(\ref{density}), it is trivial to see that 
  \begin{equation}
 \int dx \; n(x+a) \hat h_{A,1}  \approx  \int dx \;  \left[-\frac{2}{\pi^2} (\partial_x \phi)^2  \right] 
 \end{equation}
 with $\hat \phi(x+a) \approx \hat\phi(x)+ a \partial_x \hat \phi$, we have approximated $\partial_x \hat \phi(x+a) \approx \partial_x [\hat \phi(x)+ a \partial_x \hat \phi] \approx \partial_x \hat \phi $
and  ignored terms of the form $\partial_x^2 \hat\phi \partial_x \hat \phi$ and $e^{\pm 2ik_F x}$ fast oscillating terms.
Integrating by parts, we also observe that 
\begin{equation}
 \int dx \; n(x+a) \hat h_{A,3} = -  \int dx \; \left(n(x+a) \hat h_{A,2} +\hat C\right)
 \label{vioz}
\end{equation}
 with $\hat C \propto \partial_x \hat \phi \partial^2_x \hat \phi$. Then the first integral in Eq.~(\ref{vioz}) cancels with $\int dx \; n(x+a) \hat h_{A,2}$, while the $\hat C$ term can be neglected.

Finally, we consider $   \int dx \; n(x+a) \hat h_{B}$. Neglecting fast oscillating terms $e^{\pm 2ik_Fx}$, we get $   \int dx \; n(x+a) \hat h_{B} \approx 
   \int dx \; (\hat h_{\rm non-osc} + \hat h_{4k_F}) $. The term $\hat h_{\rm non-osc}$ consists of 
the following non-oscillating terms:
\begin{eqnarray}
\hat d^\dagger_R(x+a) \hat d_L(x+a) \hat d^\dagger_L(x) \hat d_R(x+2a) =  \frac{1}{(2\pi a)^2} e^{2ia \partial_x \hat \theta} \\
\hat d^\dagger_R(x+a) \hat d_L(x+a)  \hat d^\dagger_L(x+2a) \hat d_R(x)  = \frac{1}{(2\pi a)^2} e^{-2ia \partial_x  \hat \theta} 
\end{eqnarray} 
plus their hermitian conjugates; these terms can be rewritten as
\begin{equation}
  \frac{1}{(2\pi a)^2} e^{\pm 2ia \partial_x \hat \theta} \approx \frac{1}{(2\pi a)^2} \left[ 1 \pm 2ia \partial_x \hat \theta - 2a^2 (\partial_x \hat \theta)^2 \right] \,.
\end{equation}
The contribution  $\hat h_{4k_F}$ consists of the following terms 
\begin{eqnarray}
\hat d^\dagger_R(x+a) \hat d_L(x+a) \hat d^\dagger_R(x) \hat d_L(x+2a) =\frac{1}{(2\pi a)^2} e^{i\left[ 4 \hat \phi(x) +4a \partial_x \hat \phi + 2a \partial_x \hat \theta \right] }  \\
\hat d^\dagger_R(x+a) \hat d_L(x+a)  \hat d^\dagger_R(x+2a) \hat d_L(x)  = \frac{1}{(2\pi a)^2} e^{i\left[ 4 \hat \phi(x) +4a \partial_x\hat  \phi - 2a \partial_x \hat \theta \right] }  
\end{eqnarray} 
and of their hermitian conjugates; when bosonized they give sine-Gordon terms of the form $\cos 4\hat \phi$. 
Collecting the results, we have
\begin{equation}
 \hat H^{(2)}_{\rm U} = -\tilde U a \left[ -\frac{2}{\pi^2} (\partial_x \hat \phi)^2 -\frac{2}{\pi^2} (\partial_x \hat \theta)^2+  \frac{4}{(2\pi a)^2} \cos 4\hat \phi \right]  \,.
 \end{equation}
Finally, the original Hamiltonian can be recast into the form
\begin{equation}
\hat H=  \frac{u}{2\pi} \int dx \left[ \frac{1}{K}(\partial_x \hat \phi)^2 +  K(\partial_x \hat \theta)^2 \right] -\frac{8\tilde U a}{(2\pi a)^2} \int dx\; \cos 4\hat \phi  
\end{equation}
provided we identify $u/K =  v_F + 12\tilde U a/\pi $ and $uK= v_F + 4\tilde U a/\pi  $.

\section{Mean field approach}
\label{app_mf}

To calculate the critical interaction for which the system becomes gapped at filling $\nu=1/2$ we proceed in the following way. 
We use a mean field approach to treat the correlated hopping term in the effective Hamiltonian of Eq.~(\ref{HUU}) and we define  
\begin{equation}
\hat H_{\rm corr} = - \frac {\tilde U}{2}  \sum_j  \big( \hat n_j \hat d^\dagger_{j+1} \hat d_{j-1} +\mathrm{H.c.} \big)\,.
\end{equation} 
Using the continuum limit operators $\hat d(x)$ and $\hat d^\dagger(x)$, we have:
\begin{equation}
\hat H_{\rm corr} \propto \int dx \;  \hat n(x+a) \left[\hat d^\dagger(x+2a) \hat d(x) + \mathrm{H.c.} \right] \,.
\end{equation}
Then, we define $\hat n(x)= n+ \delta \hat n(x)$ and $\hat d^\dagger(x+2a) \hat d(x) = \chi + \delta \hat \chi (x)$, with 
$n(x) = \langle \hat n(x) \rangle$ and $\chi (x) = \langle \hat d^\dagger(x+2a) \hat d(x)  \rangle$ such that
 \begin{equation}
\hat H_{\rm corr} \propto \int dx \;  \left[ n+\delta \hat n(x+a) \right] \left[\chi + \delta \hat \chi(x) + \mathrm{H.c.} \right].
\end{equation}
Neglecting quadratic fluctuation terms, trivial constants and an overall chemical potential, we obtain
 \begin{equation}
\hat H_{\rm corr} \propto n \int dx \;    \left[\delta \hat \chi(x) + \mathrm{H.c.}\right] \
\end{equation}
 if we approximate $\hat d(x+2a) \approx \hat d(x) +2a\partial_x \hat d(x)$, it is trivial to see that $\hat H_{\rm corr}=0$ (see also App. \ref{app_bos}).


\end{document}